\PassOptionsToPackage{table}{xcolor}
\documentclass[sigconf]{acmart}

\copyrightyear{2026}
\acmYear{2026}
\setcopyright{cc}
\setcctype{by}
\acmConference[CHI EA '26]{Extended Abstracts of the 2026 CHI Conference on Human Factors in Computing Systems}{April 13--17, 2026}{Barcelona, Spain}
\acmBooktitle{Extended Abstracts of the 2026 CHI Conference on Human Factors in Computing Systems (CHI EA '26), April 13--17, 2026, Barcelona, Spain}
\acmDOI{10.1145/3772363.3798463}
\acmISBN{979-8-4007-2281-3/2026/04}

\usepackage{cleveref}
\graphicspath{ {figures/} }

\begin{document}

\title[]{Effects of Generative AI Errors on User Reliance Across Task Difficulty}

\author{Jacy Reese Anthis}
\authornote{Work done at Microsoft Research}
\orcid{0000-0002-4684-348X}
\affiliation{
  \institution{Stanford University}
  \city{Stanford}
  \state{California}
  \country{USA}
}
\affiliation{
  \institution{University of Chicago}
  \city{Chicago}
  \state{Illinois}
  \country{USA}
}

\author{Hannah Cha}
\authornotemark[1]
\orcid{0009-0002-1476-9518}
\affiliation{
  \institution{Stanford University}
  \city{Stanford}
  \state{California}
  \country{USA}
}

\author{Solon Barocas}
\orcid{0000-0003-4577-466X}
\affiliation{
  \institution{Microsoft Research}
  \city{New York}
  \state{New York}
  \country{USA}
}

\author{Alexandra Chouldechova}
\orcid{0000-0002-2337-9610}
\affiliation{
  \institution{Microsoft Research}
  \city{New York}
  \state{New York}
  \country{USA}
}

\author{Jake Hofman}
\orcid{0000-0002-9364-9604}
\affiliation{
  \institution{Microsoft Research}
  \city{New York}
  \state{New York}
  \country{USA}
}

\begin{abstract}
    The capabilities of artificial intelligence (AI) lie along a jagged frontier, where AI systems surprisingly fail on tasks that humans find easy and succeed on tasks that humans find hard. To investigate user reactions to this phenomenon, we developed an incentive-compatible experimental methodology based on diagram generation tasks, in which we induce errors in generative AI output and test effects on user reliance. We demonstrate the interface in a preregistered 3x2 experiment ($N$~=~577) with error rates of 10\%, 30\%, or 50\% on easier or harder diagram generation tasks. We confirmed that observing more errors reduces use, but we unexpectedly found that easy-task errors did not significantly reduce use more than hard-task errors, suggesting that people are not averse to jaggedness in this experimental setting. We encourage future work that varies task difficulty at the same time as other features of AI errors, such as whether the jagged error patterns are easily learned.
\end{abstract}



\maketitle

\section{Introduction}
\label{sec:introduction}

The general-purpose capabilities of artificial intelligence (AI), particularly generative AI tools such as large language models (LLMs), create significant cognitive challenges for users~\cite{sturgeon_humanagencybench_2025, subramonyam_bridging_2024, tankelevitch_metacognitive_2024, zamfirescu-pereira_why_2023}. Effective use requires forming and updating accurate expectations of the tool's effectiveness across diverse tasks. Accurate calibration is made challenging by the complexity and opacity of the tool~\cite{zhao_explainability_2024}, diversity of use cases~\cite{raiaan_review_2024}, and highly personalized experiences that emerge over long periods of use~\cite{manoli_shes_2025}. This is challenging for users, developers who aim to build useful tools, and researchers aiming to understand tool development and use.

A crucial challenge humans face is accounting for the ``jagged frontier''~\cite{dellacqua_navigating_2023, gans_model_2026, morris_characterizing_2026, zhou_larger_2024}, the fact that modern AI systems are difficult to predict because they can fail on tasks that humans find easy despite superhuman performance on tasks that humans find hard. For example, AI systems can fail to count letters in common words or answer riddles correctly when rephrased~\cite{mccoy_embers_2024, west_generative_2023, zhou_larger_2024}, but they can also summarize vast amounts of text~\cite{skarlinski_language_2024} and speak dozens of languages~\cite{zhao_how_2024}. Even if a human observes an example of AI performance, they are often not able to effectively update their predictions of AI performance on a different task~\cite{vafa_large_2024}.

To understand human-AI interaction in light of such challenges, we developed an incentive-compatible experimental methodology to study human reactions to generative AI errors that balances realism, experimental control, and the ability to quantify user reliance. Our approach involves generating diagrams of the type used in project planning~\cite{google_workspace_ai_nodate} and slideshow presentations~\cite{lucidchart_diagramming_nodate}. In Phase 1, participants learn about the AI tool's performance by observing a series of tasks in which they are shown a prompt given to the tool and the resulting diagram, with the pattern of successes and errors determined by random assignment to treatment groups. In Phase 2, participants apply what they have learned about the AI tool in a series of willingness‑to‑pay (WTP) tasks~\cite{becker_measuring_1964} as a measure of reliance: in each task, they can pay to use the AI tool to attempt the task, and they receive a monetary reward if the AI tool successfully generates the intended diagram. This way, we can observe effects of error patterns in Phase 1 on user reliance in Phase 2.

Using this methodology, we ran a preregistered 3x2 experiment ($N$~=~577) in which participants were randomly assigned to one of six Phase 1 treatments: observing an error rate of 10\%, 30\%, or 50\%, with those errors being on either the hard tasks or the easy tasks. In this context, ``easy'' and ``hard'' refer to the complexity of the diagram that must be generated, where a human would find it easier to quickly create a simpler diagram. As expected, a larger number of errors led to lower WTP. However, contrary to expectations, we did not find a consistent effect of task difficulty. These findings raise important questions about how humans will under- or over-rely on AI tools and how human-AI collaboration will evolve as AI tools become more powerful. We propose follow-up research that further investigates human reactions by varying not just task difficulty but also how easy the error pattern is for humans to learn. People may be comfortable with non-humanlike AI systems as long as the systems are predictable, or they may prefer humanlike systems even if the systems are not more predictable.

\section{Background}
\label{sec:background}

People stop using relatively simple algorithmic tools after seeing even minor mistakes, penalizing the tool more harshly than they would a human advisor, a pattern termed algorithm aversion~\cite{dietvorst_algorithm_2015}, yet others have identified cases of automation bias~\cite{goddard_automation_2012} and algorithm appreciation~\cite{castelo_task-dependent_2019, you_algorithmic_2022} in which people favor algorithms over human judgment. These patterns can lead to underreliance and overreliance, where humans fail to correctly identify errors~\cite{bansal_beyond_2019}. Human failures to respond to errors appropriately are a contributing factor in the failure of human–AI teams to attain ``strong synergy,'' an ideal interaction in which teams outperform humans and AI systems that complete tasks individually~\cite{vaccaro_when_2024}.

AI reliance depends on various factors, such as whether tasks are perceived as objective or subjective~\cite{castelo_task-dependent_2019}, whether humans merely provide input to the algorithm versus being able to override the algorithmic outcome~\cite{cheng_overcoming_2023}, and whether the algorithm is described as a novice or an expert~\cite{khadpe_conceptual_2020}. Appropriate reliance can be facilitated through system design, including explanations of the algorithmic decision-making process~\cite{schoeffer_explanations_2024, vasconcelos_explanations_2023} and encouragement for users to reflect thoughtfully on whether to rely on the algorithm~\cite{bucinca_trust_2021}.

However, much of this evidence is based on interaction with conventional tools. For example, \citet{bansal_updates_2019} tested advice that consisted of a binary choice (e.g., whether a product was defective), and \citet{dhuliawala_diachronic_2023} mapped trust recovery following errors in which the system made a simple multiple-choice recommendation on a math or trivia problem. Some recent work has moved into generative settings, such as highlighting parts of text or code where the system is particularly uncertain~\cite{spatharioti_effects_2025, vasconcelos_generation_2024}, but empirically grounded accounts of what drives trust and reliance in generative workflows remain unsettled. Modern workflows often take open-ended inputs in natural language~\cite{subramonyam_bridging_2024}; involve repeated interaction with agentic systems that affect the user's own agency~\cite{sturgeon_humanagencybench_2025, wang_towards_2021}; and involve the manipulation of large-scale artifacts, such as documents and codebases~\cite{feng_cocoa_2025}. It is unclear to what extent work on conventional algorithms extends to modern human-AI interaction.

An AI system with a jagged frontier of capabilities, performing poorly on tasks humans find easy and well on tasks humans find hard, is distinctly non-humanlike. Humanlikeness is known to be consequential. In empirical studies, human–computer interaction researchers have found significant effects of the humanlikeness of robots~\cite{roesler_meta-analysis_2021}, synthetic voices~\cite{kuhne_human_2020, kim_human_2022}, and visual avatars~\cite{green_sensitivity_2008, zhang_examining_2024} on human expectations and behavior. There are reasons why users could rationally prefer humanlike or non-humanlike tools. Humanlikeness could be beneficial by making the system easier to predict and use: allowing the user to quickly leverage interaction patterns they know from human–human interaction. On the other hand, humanlikeness could reduce efficiency by reducing complementarity: for example, if a human and a humanlike AI have the same factual knowledge and recall ability, then the human–AI team may not do any better on a knowledge-based test than either the human or AI would alone. Rational-actor models may also fail to predict user behavior, particularly because generative AI introduces substantial metacognitive demands that make reasoning difficult~\cite{tankelevitch_metacognitive_2024}.

The unique features of generative AI could limit the applicability of classical human-computer interaction theory and studies of conventional algorithmic tools. Classically, computer users must bridge two major gulfs: the \textit{gulf of execution}, in which they get the computer to execute a task, and the \textit{gulf of evaluation}, in which they work to understand what the computer did and ensure its correctness~\cite{norman_design_1988}. Recently, \citet{subramonyam_bridging_2024} proposed a fundamental extension for generative AI by adding the \textit{gulf of envisioning}, in which the user must formulate an input (i.e., prompt) that accounts for the tool's flexibility, ambiguity in the user's intent, and indeterminacy of the output. It is often challenging for users to correctly steer a generative AI model into correctly producing a particular output, such as writing a text prompt for a diffusion model to reproduce an image shown to the user~\cite{vafa_whats_2025}, and users must navigate significant uncertainty as to whether the user or tool is more at fault when an error is made~\cite{jahani_prompt_2026, schoenherr_when_2024}. Our study aims to extend the study of errors in conventional algorithmic decision aids to these challenges.

\section{Hypotheses}
\label{sec:hypotheses}

We preregistered the following hypotheses, in which $\mu$ denotes the mean bid of participants, such as $\mu_{1}$ for participants who saw exactly one error and $\mu_{\text{easy}}$ for participants who saw errors on easy tasks. First, we expected that an increase in the number of errors would reduce AI use.

Second, we hypothesized that errors on easier tasks would reduce AI use more than errors on harder tasks because easy-task errors are more surprising and increase user uncertainty about the tool's performance. Prior work by \citet{papenmeier_how_2022} found that people viewed a predictive model as less accurate when it made errors on tasks described as ``easy'' rather than ``difficult'' or ``impossible,'' and \citet{raux_human_2025} similarly found that people have lower accuracy estimates when an error is made on an ``easy'' rather than a ``hard'' task.

Third, we considered an interaction effect. Because we expected the presence of multiple errors to be more salient to participants than a single error, we expected this to increase the effect that task difficulty would have on AI use.

Finally, we posed two hypotheses to probe the relative impact of the number of errors and the task difficulty on AI use. These relate to the intuition that making easy-task errors rather than hard-task errors is ``at least as bad'' as making two additional errors, either from one to three or from three to five. A hypothesis test for this requires a margin, which we set at \$0.05 based on pilot data and a practical sense of what effect size would be meaningful in this context.

\begin{figure*}[t]
\centering
\fbox{
  \begin{minipage}{0.98\textwidth}
    \vspace{6pt}
    \centering
    \vbox{
      \halign{#\hfil\cr
          \textbf{H1}: An AI tool making more errors reduces AI use: $ (\mu_{3} < \mu_{1}) \land (\mu_{5} < \mu_{3}) $ \cr
          \noalign{\vskip 4pt}
          \textbf{H2}: Easy-task AI errors reduce AI use more than hard-task errors: $ \mu_{\text{easy}} < \mu_{\text{hard}} $ \cr
          \noalign{\vskip 4pt}
          \textbf{H3}: Easy-task AI errors reduce AI use more than hard-task errors even more when there are multiple errors: \cr
          \hfil $ \mu_{(3,5),\text{hard}} - \mu_{(3,5),\text{easy}} > \mu_{1,\text{hard}} - \mu_{1,\text{easy}} $ \hfil\cr
          \noalign{\vskip 4pt}
          \textbf{H4}: Easy-task AI errors reduce AI use at least as much as two additional hard-task AI errors: \cr
          \hfil \textbf{(a)} $ \mu_{3,\text{easy}} - \mu_{5,\text{hard}} < \$0.05 $; \textbf{(b)} $ \mu_{1,\text{easy}} - \mu_{3,\text{hard}} < \$0.05 $ \hfil \cr
      }
    }
    \vspace{6pt}
  \end{minipage}
}
\end{figure*}

\section{Methodology}
\label{sec:methodology}

We built an interface for a diagram generation task, which we designed to be understandable to participants, expressive for a variety of practical contexts, amenable to procedurally inducing errors, completable with nontrivial effort by a human or current LLM, and based on natural language input. We leveraged this interface for a preregistered experiment (\url{https://aspredicted.org/p7ge92.pdf}) and have shared the data and reproducible analysis code online (\url{https://github.com/jacyanthis/ai-errors-chi-ea-2026}). The study received institutional ethics approval prior to data collection, and participants gave their informed consent and were debriefed after completing the study.

\begin{figure*}[ht]
    \centering
    \includegraphics[width=\linewidth]{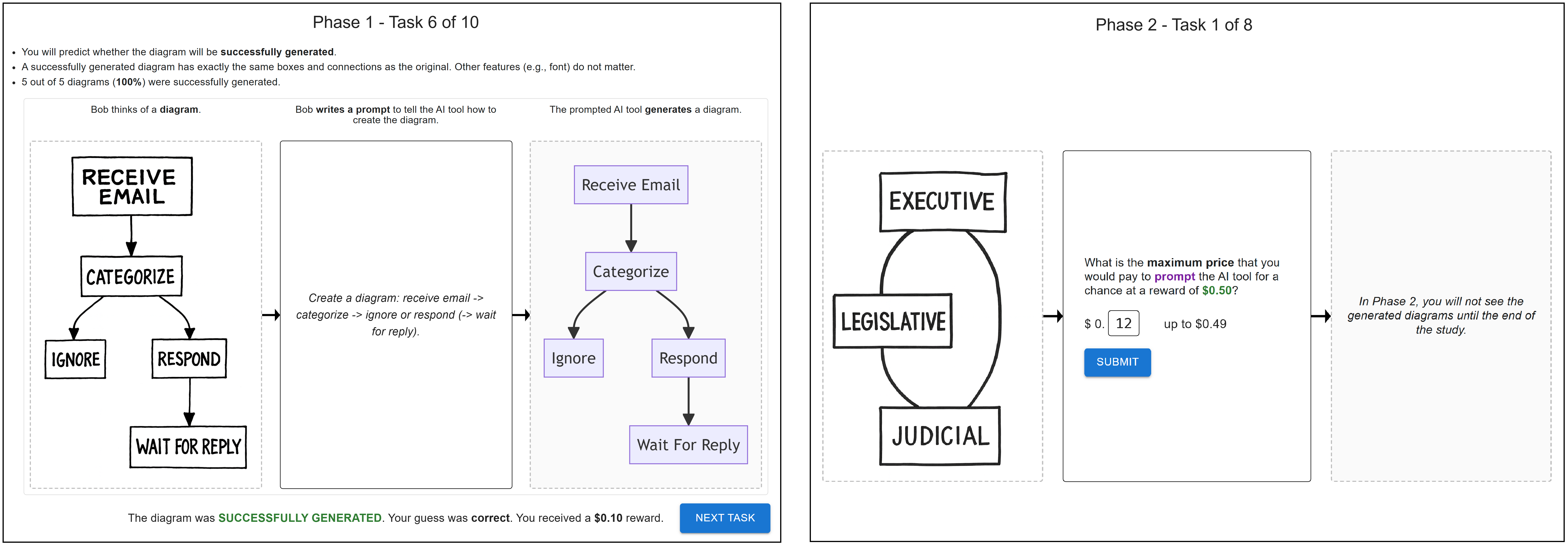}
    \caption{Screenshots of the study interface. In Phase 1 (left), participants predict whether the AI tool will successfully generate the diagram. In Phase 2 (right), participants report their WTP for an opportunity to use the AI tool for a monetary reward.}
    \Description{Two screenshots of the study interface are shown side-by-side. In Phase 1 (left), participants see a sketch of the target diagram and prompt, and they predict whether the AI tool will successfully generate the diagram. This participant correctly predicted that the diagram would be successfully generated. In Phase 2 (right), participants report their WTP for an opportunity to use the AI tool for a monetary reward. This participant typed in 12 cents for a relatively simple diagram.}
    \label{fig:methodology}
\end{figure*}

We developed the study interface as a React app. Participants were introduced to the task and completed the demonstration phase (Phase 1), in which they observed a sketch shown to a hypothetical person and a text prompt written by that hypothetical person. To facilitate engagement, participants were asked to predict whether the AI tool would successfully generate the diagram, earning \$0.10 if they guessed correctly (\Cref{fig:methodology}, left). Afterward, they viewed a summary of the Phase 1 results (e.g., the percentage of errors made) and then were introduced to the WTP bidding system and completed the measurement phase (\Cref{fig:methodology}, right), as described in the following section. Participants were not shown the result of their text prompts until the end of the study in order to prevent learning effects.

We recruited a sample of U.S. adults from Prolific, using the platform's representative quota-based sampling across age, gender, and race. We conducted an attention check, which 93.5\% of participants passed, resulting in a final sample of 577. Pilot data suggested that this sample size would provide more than 80\% power for detecting a 0.3 standardized effect size (approximately \$0.05) for each of the hypotheses. Throughout the study, we monitored time spent, clicks, and other interaction data, and we blocked pasting of text into the study interface to mitigate the usage of automated tools, but we found no participants for whom data exclusion seemed warranted.

Participants were randomly assigned to one of six conditions: out of the ten task completions in the demonstration phase, participants saw either one, three, or five errors, and these errors were either in the tasks to generate the simplest diagrams (i.e., easy tasks) or the most complex diagrams (i.e., hard tasks). Odd-numbered tasks were to generate linear diagrams of three, four, five, six, or seven nodes, and even-numbered tasks were to generate diagrams with the same number of nodes but with one non-linearity (e.g., a fork in which Node A is connected to both Node B and Node C). Task ordering was randomized between first-to-tenth-task (i.e., easiest to hardest, approximately) and tenth-to-first-task (i.e., hardest to easiest, approximately); other orderings were avoided to make it easier for participants to detect patterns.

\subsection{Measures}

\textit{Willingness to pay.} Participants entered a bid to indicate their WTP for the opportunity to use an AI tool and, if successful in the task, receive a \$0.50 reward, an incentive-compatible procedure known as the Becker–DeGroot–Marschak method~\cite{becker_measuring_1964}. Participants paid this money out of a fixed endowment that was equal across all participants, regardless of Phase 1 outcome. For each Phase 2 task, a random price is drawn, and the participant pays the price and completes the task if their bid is at least as high as the price. For example, if a user is willing to pay \$0.25, that means they would pay any price up to \$0.25, which implies they believe the probability of receiving the \$0.50 is sufficiently high to be worth paying \$0.25 upfront, including the possibility that they lose money without any reward. If a user is only willing to pay \$0.00, then they have no chance of using the AI tool. The maximum bid was fixed below the reward amount because a bid of the reward amount would have no potential gain over a bid of one cent fewer. Participants undertook a brief tutorial to introduce them to this method, and no participants described being confused by the procedure in the debrief.

\textit{Expected difficulty.} For each measurement phase task, we first asked participants ``How difficult does this task seem?'' based on the sketch with answer choices of: ``It would take me one attempt,'' ``It would take me two attempts,'' ``It would take me a few (3–5) attempts,'' ``It would take me many (6+) attempts,'' ``I think the AI tool can do this, but I don't think I could figure out the prompt,'' and, ``I think the AI tool cannot do this with any prompt.''

\textit{Performance expectancy and effort expectancy.} After providing informed consent and before being introduced to the task, participants responded to two brief indices based on \citet{venkatesh_consumer_2012}: four items based on their expected performance with AI (e.g., ``Using AI increases my productivity'') and three items based on the expected effort they believe it takes them to use AI (e.g., ``I find AI easy to use''). Participants responded to the same indices after the measurement phase, providing a supplementary pre-post survey measure.

\textit{Personal characteristics.} After the post-task survey measures, participants were asked four questions regarding AI use (e.g., ``How often do you interact with AI?''), followed by a variety of demographic questions (political leaning, age, gender, race/ethnicity, education level, household income, religion).

\section{Results}
\label{sec:results}

\begin{figure}[t]
    \includegraphics[width=\linewidth]{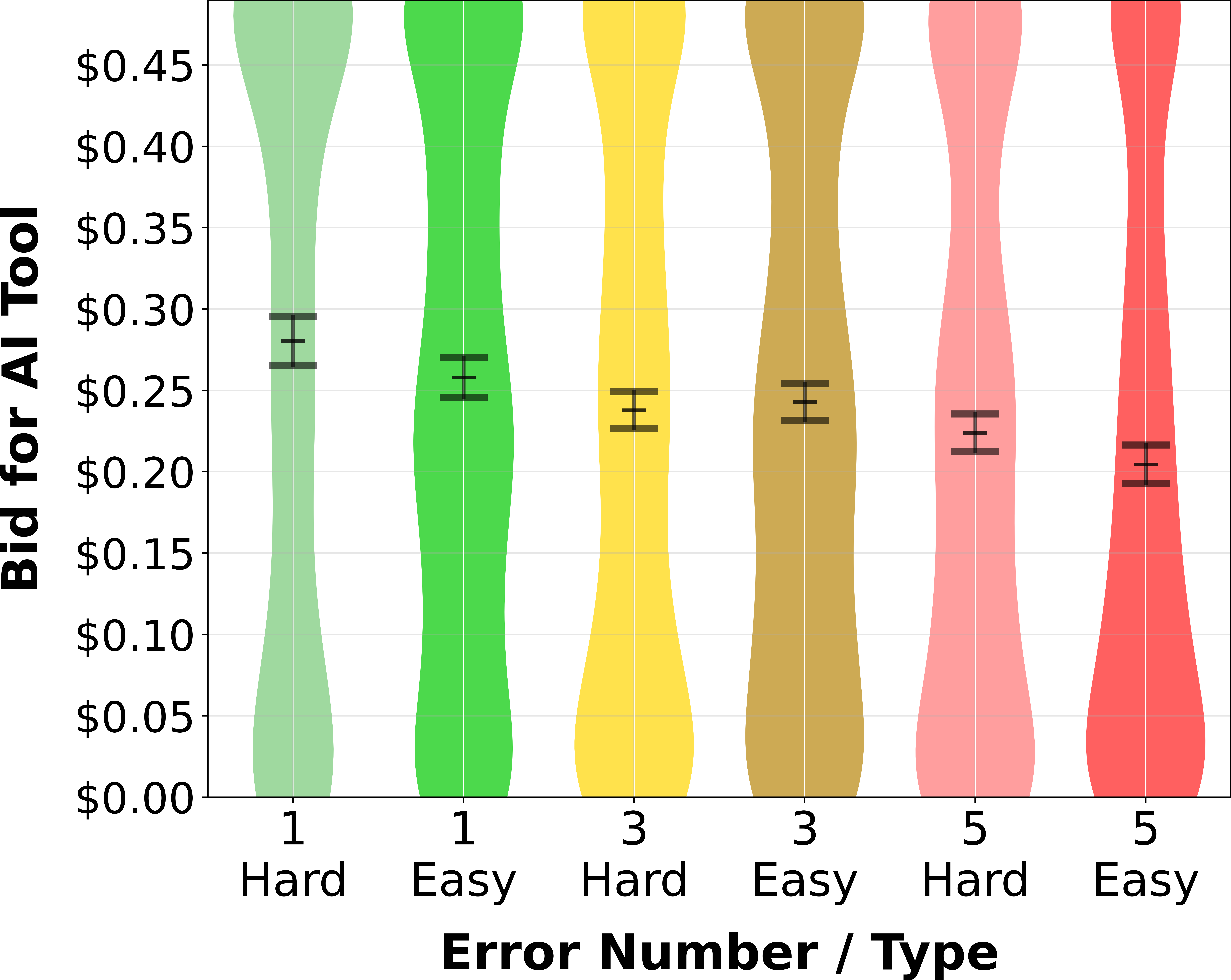}
    \caption{Bids across the six experimental conditions. Means are model-adjusted, and error bars show standard errors.}
    \Description{The violin plot shows distributions for each of the six experimental conditions.}
    \label{fig:results}
\end{figure}

Hypotheses were tested with a mixed-effects linear model that predicted WTP based on number of errors, difficulty of tasks on which errors were made, randomization factors, pre-task survey measures, and demographics.

\textit{H1: Errors reduce AI use.} The results supported H1 with a mean bid of \$0.27 for 1-error participants, \$0.24 for 3-error participants, and \$0.21 for 5-error participants. This resulted in $\mu_1 - \mu_3 > 0$ with a difference of \$0.03 (SE = \$0.01, $t$ = 2.36, $p$ = 0.009) and $\mu_3 - \mu_5 > 0$ with a difference of \$0.02 (SE = \$0.01, $t$ = 1.92, $p$ = 0.028).

\textit{H2: Easy-task errors reduce AI use more than hard-task errors.} The results did not support H2. We found a mean bid of \$0.23 for participants exposed to easy-task errors and \$0.25 for participants exposed to hard-task errors (diff = \$0.01, SE = \$0.01, $t$ = 0.990, $p$ = 0.161). Exploratory follow-up analyses (\Cref{sec:prior_AI}) suggest there may be a subgroup-specific effect of task difficulty on participants who rarely or never consume AI-related content. Notably, in the analogous hypothesis test for a different outcome, the pre-post difference in performance expectancy, there is a significant difference (diff = 0.09, SE = 0.05, $t$ = 1.803, $p$ = 0.036).

\textit{H3: Easy-task errors reduce AI use more than hard-task errors even more when there are multiple errors.} The results did not support H3. Between 1-error/easy-task participants and 1-error/hard-task participants, we found a difference of \$0.02, and between multiple-error/hard-task participants and multiple-error/easy-task participants, we found a difference of \$0.01. The difference-in-differences was -\$0.01 (SE = \$0.03, $t$ = -0.62, $p$ = 0.733).

\textit{H4: Easy-task AI errors reduce AI use at least as much as two additional hard-task AI errors.} The results supported H4a with a difference of bids between 3-error/easy-task and 5-error/hard-task participants that was significantly less than the preregistered margin of \$0.05, at -\$0.02 (SE = \$0.02, $t$ = -2.14, $p$ = 0.016). The results did not, however, support H4b. Between bids of 1-error/easy-task and 3-error/hard-task participants, we found a difference that was not significantly less than the preregistered margin of \$0.05, at \$0.02 (SE = \$0.02, $t$ = -1.64, $p$ = 0.050).

\section{Discussion}
\label{sec:discussion}

Our findings confirmed that observing more errors reduces AI use. However, contrary to our expectations, whether the AI tool made errors on easier or harder tasks did not significantly affect user reliance, even though we found evidence of an effect on pre-post performance expectancy and evidence of a subgroup-specific effect for participants with low AI-content consumption. In this experiment, we set out to manipulate the task difficulty on which AI errors are made by creating conditions that either aligned with intuitive error patterns (i.e., hard-task errors) or ran counter to those prior beliefs (i.e., easy-task errors). We hypothesized that easy-task errors would reduce use more than hard-task errors by reducing the perceived predictability of the tool, based on the assumption that participants would have less confidence in their ability to accurately predict the AI tool's performance when its performance did not track the ordering of task difficulty set by human performance; therefore, risk aversion would lead participants to bid less to use the tool with easy-task errors than with hard-task errors.

To explain our results and motivate future work, we propose that human reactions to the ``jagged frontier'' depend on two distinct factors: \textit{alignment} of the frontier with the user's prior expectations and \textit{clarity} with which the user perceives the frontier, meaning the certainty they have in their estimates of the tool's capabilities. In this framework, reliance is affected by the expected performance of the tool and the clarity of the frontier, and alignment with prior expectations affects how easily clarity is achieved: When a tool's behavior aligns with prior expectations, users can generalize from fewer observations. Clarity can also be achieved without alignment through sufficient experience or through tool behavior that is highly salient (i.e., easy to notice and understand).

Neither alignment nor clarity requires humanlikeness. For example, calculators are very unlike humans (e.g., superhuman at arithmetic but incapable of language or movement), yet their behavior is well-aligned with our expectations, and users rely on them without hesitation. Prior work suggests interacting with AI systems through social interaction, such as conversational interfaces and natural language, increases the likelihood that users perceive these systems as humanlike~\cite{klein_effects_2025}. Thus, in the context of conversational AI systems, humanlikeness can be viewed as the alignment between the perceived frontier of tool capabilities and the frontier of human performance on the same task. In our experiment, easy-task errors were less humanlike, which can affect user reliance alongside alignment and clarity.

Easy-task and hard-task errors were similar in salience—both being results of particular task difficulty—and therefore, we expect, similar in the clarity of the jagged frontier. The differences in alignment with expectations and in humanlikeness did not result in a significant experimental difference between the task difficulty conditions. Our exploratory analyses found that the effect of task difficulty was concentrated among participants who rarely consume AI-related content (\Cref{sec:prior_AI}), suggesting that misalignment or non-humanlikeness still shape perceptions of people with weaker prior expectations of AI performance. The null result could also be explained in part by the WTP bidding methodology, given that task difficulty did significantly affect pre-post performance expectancy. In other words, alignment and humanlikeness may shape beliefs about a tool even when it does not translate into differences in willingness to pay.

\paragraph{Future Work}

To disentangle the effects of different features of the jagged frontier, we propose follow-up work that varies the saliency of error patterns. For example, errors could be made hard to detect by being associated with less salient features of the input, such as the presence of unusual characters in diagram labels, or even by being randomly distributed. An experimental cross of saliency conditions and the task difficulty of the study described above would help identify whether reliance is affected by alignment, clarity, or other features. Our expectation is that low saliency—and the resulting lack of clarity—would reduce reliance more than the misaligned error pattern. Some participants could also receive explicit information about the tool's error patterns to further increase clarity. This work could also include additional outcome measures, especially those that relate specifically to alignment or clarity.

In addition to this follow-up study, we see several avenues for future investigation to better understand reliance and trust in light of the unique metacognitive challenges of generative AI~\cite{tankelevitch_metacognitive_2024}. These avenues include studying how error patterns transfer across task domains (e.g., diagram generation, writing, mathematics, coding); the interaction between error patterns and willingness to interactively experiment and learn with experience; and teasing out the effects on perceptions of producibility—whether a tool can complete a task—and steerability—whether the user can efficiently prompt the tool to succeed~\cite{vafa_whats_2025}. Our experiment raises more questions than it answers, but it provides an entry point to understanding the different forms of jaggedness that shape human reactions.

\begin{acks}
We are grateful for input and feedback from Microsoft Research colleagues, including those in Fairness, Accountability, Transparency, and Ethics (FATE) and Computational Social Science (CSS).
\end{acks}

\bibliographystyle{ACM-Reference-Format}
\bibliography{references/anthis_zotero}

\newpage

\appendix
\section{Appendix}
\label{sec:appendix}

\subsection{Easy vs. Hard Task Examples}
\label{sec:examples}

\begin{figure*}[ht]
    \includegraphics[width=0.7\linewidth]{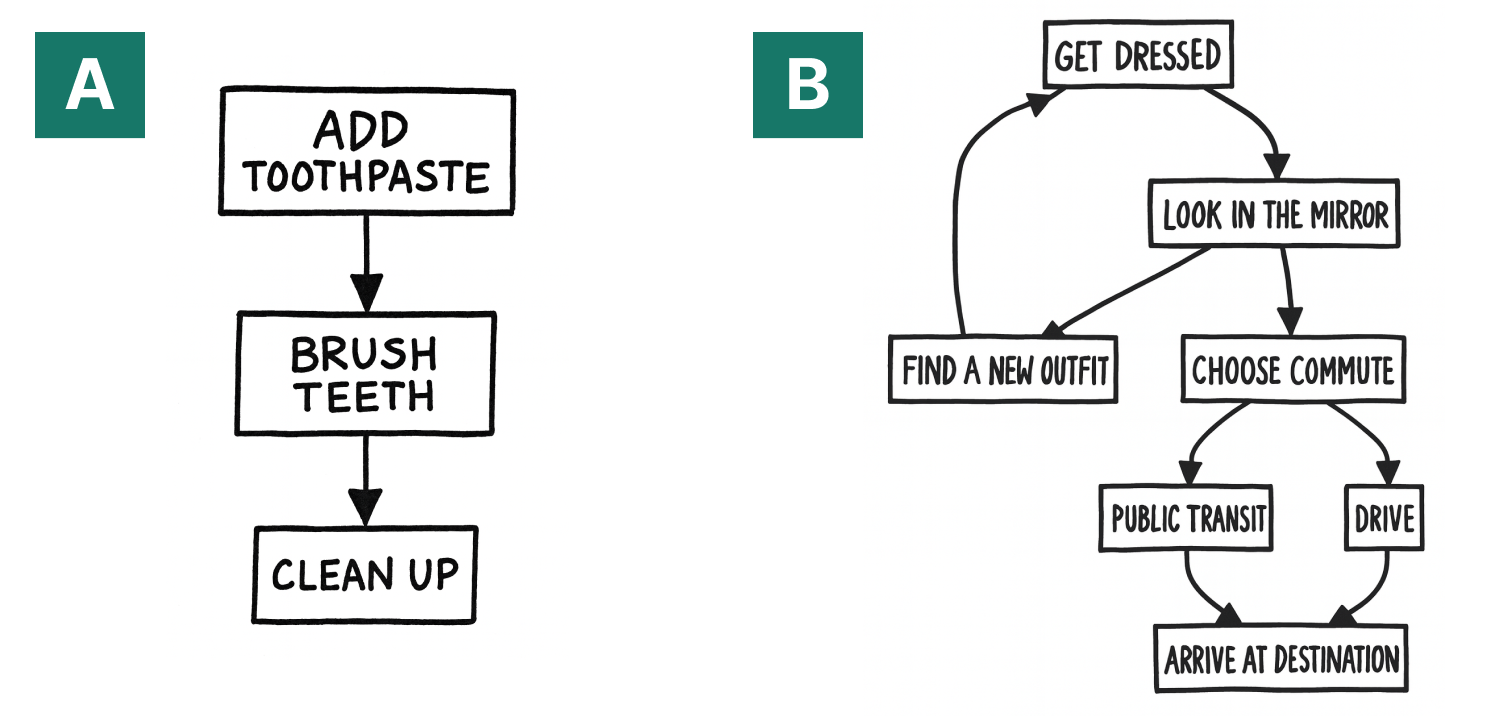}
    \caption{The easiest (i.e., simplest) (A) and hardest (i.e., most complex) (B) diagrams that participants were asked to recreate in Phase 2.}
    \Description{The figure shows an example of an easy task on the left labeled as "A" and a hard task on the right labeled as "B".}
    \label{fig:easy-v-hard}
\end{figure*}

\Cref{fig:easy-v-hard} depicts the easiest and hardest diagrams that participants were asked to recreate in Phase 2. Easy diagrams had few components with a simple linear flow, whereas harder diagrams had more nodes and non-linear connections such as forks or cycles.

\subsection{Survey Interface Overview}
\label{sec:survey_interface}

\begin{figure*}[ht]
    \centering
    \includegraphics[width=\linewidth]{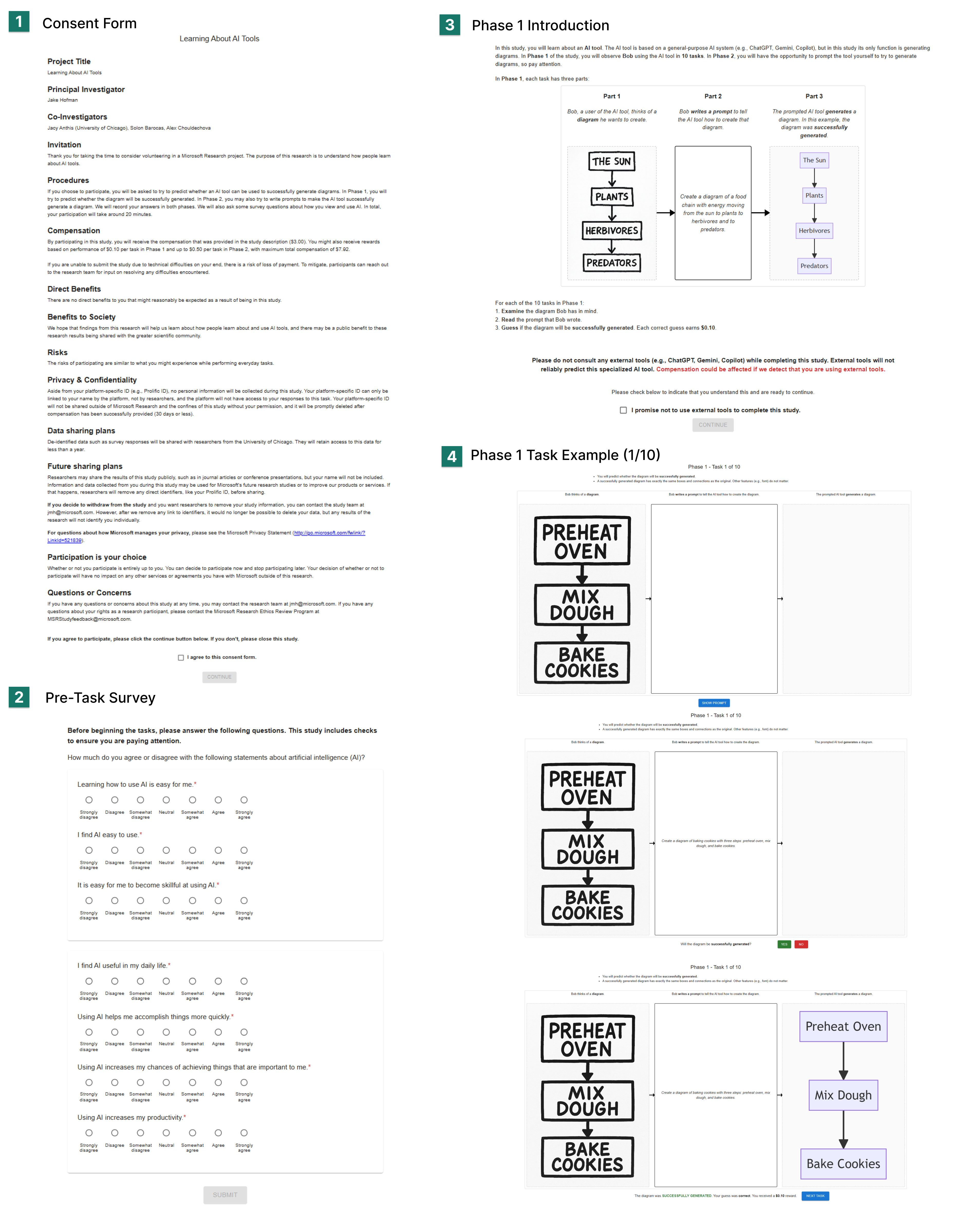}
    \caption{First stages of the study interface. This participant was randomly assigned to view the Phase 1 tasks in ascending order of difficulty (easier to harder) and to see errors later in the phase (for the more difficult tasks, not shown here). The Phase 2 task order is fully randomized.}
    \label{fig:survey-pt1}
\end{figure*}

\begin{figure*}[ht]
    \centering
    \includegraphics[width=\linewidth]{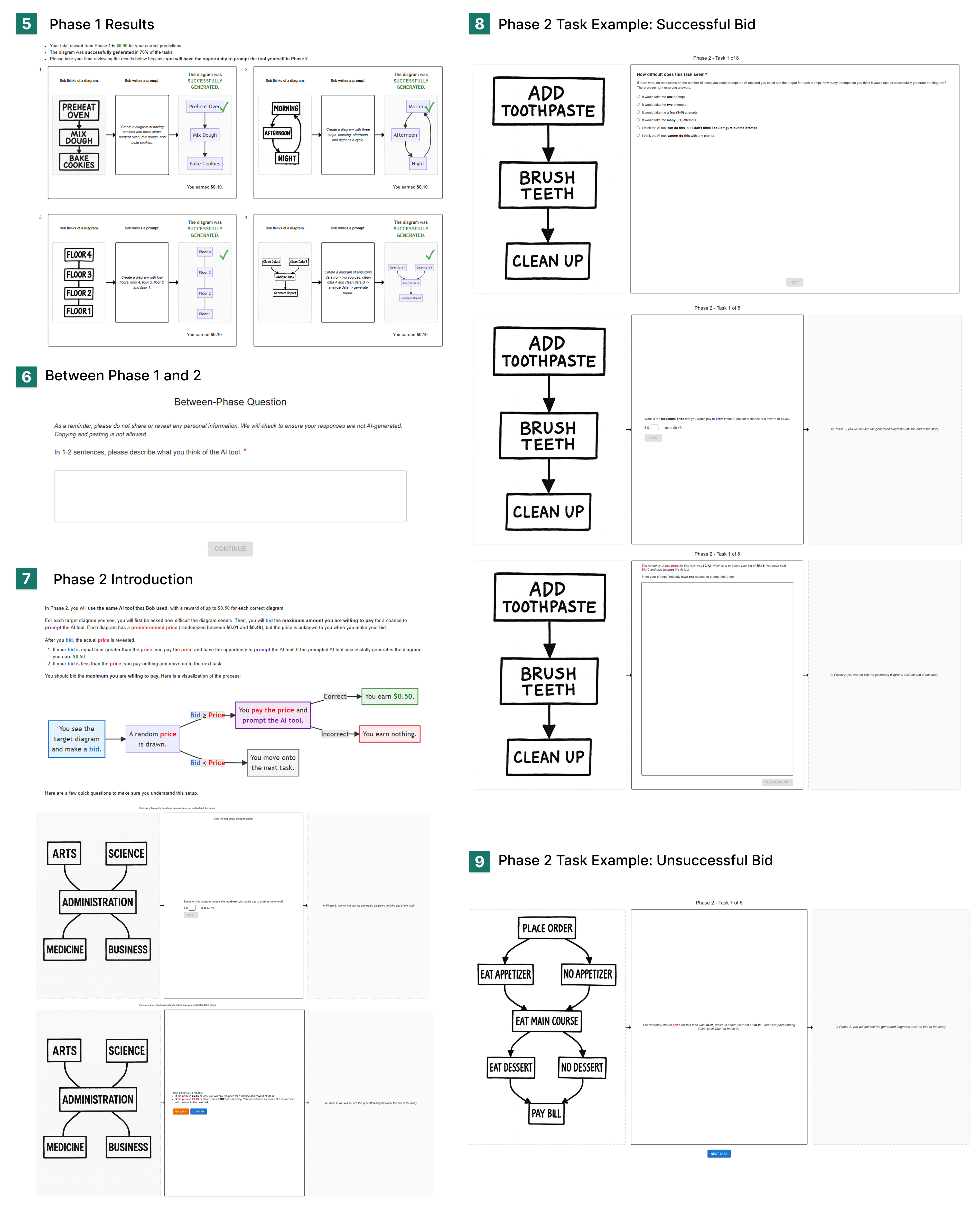}
    \caption{Additional stages of the study interface. This participant was randomly assigned to view the Phase 1 tasks in ascending order of difficulty (easier to harder) and to see errors later in the phase (for the more difficult tasks, not shown here). The Phase 2 task order is fully randomized.}
    \label{fig:survey-pt2}
\end{figure*}

\begin{figure*}[ht]
    \centering
    \includegraphics[width=\linewidth]{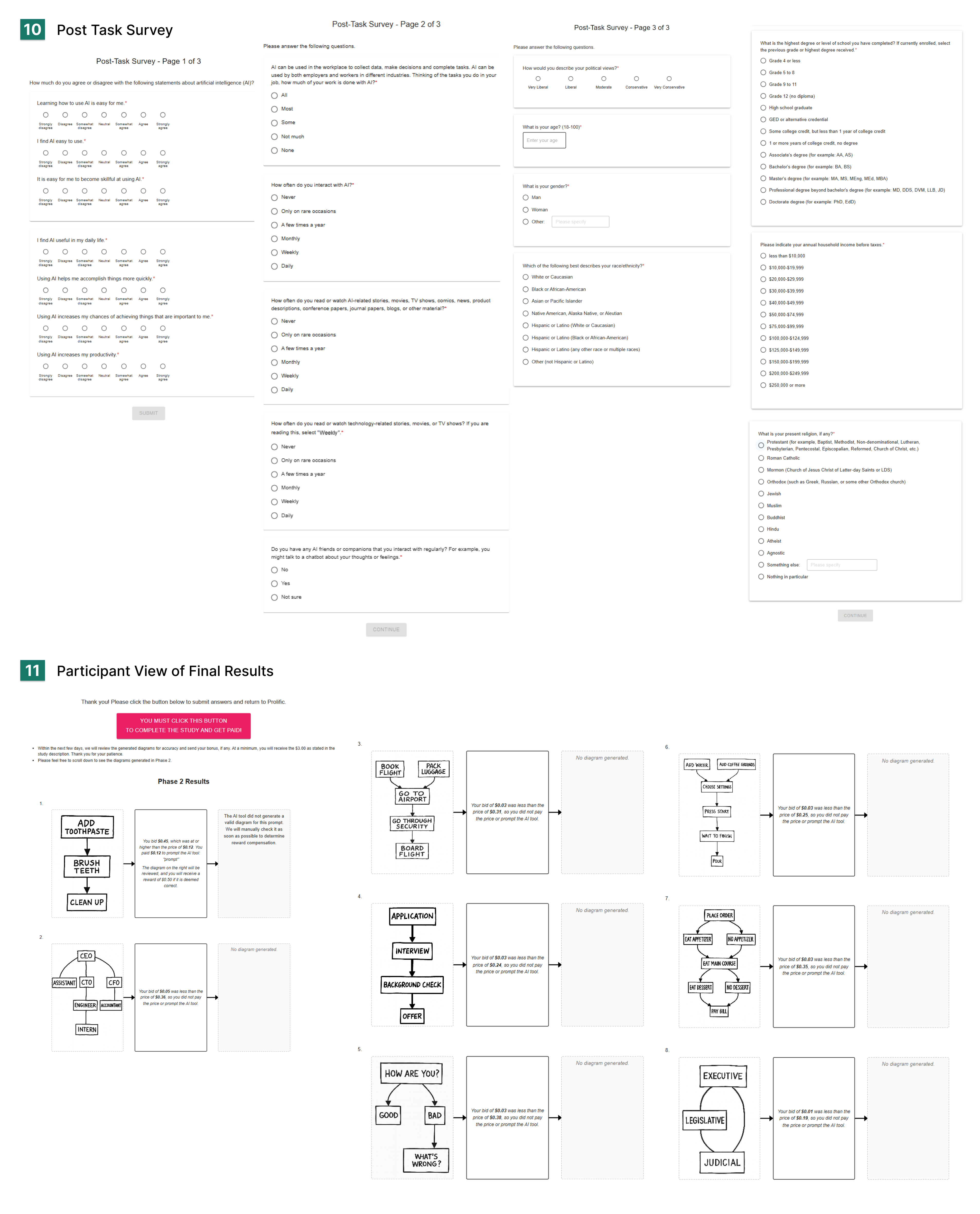}
    \caption{Additional stages of the study interface. This participant was randomly assigned to view the Phase 1 tasks in ascending order of difficulty (easier to harder) and to see errors later in the phase (for the more difficult tasks, not shown here). The Phase 2 task order is fully randomized.}
    \label{fig:survey-pt3}
\end{figure*}

\Cref{fig:survey-pt1,fig:survey-pt2,fig:survey-pt3} show the complete study interface, including the introduction, Phase 1 demonstration tasks, summary of Phase 1 results, Phase 2 WTP bidding tasks, and post-task survey measures.


\subsection{Perceived Task Difficulty By Participants}
\label{sec:perceived_task_difficulty}

\begin{figure*}[ht]
    \centering
    \includegraphics[width=0.7\linewidth]{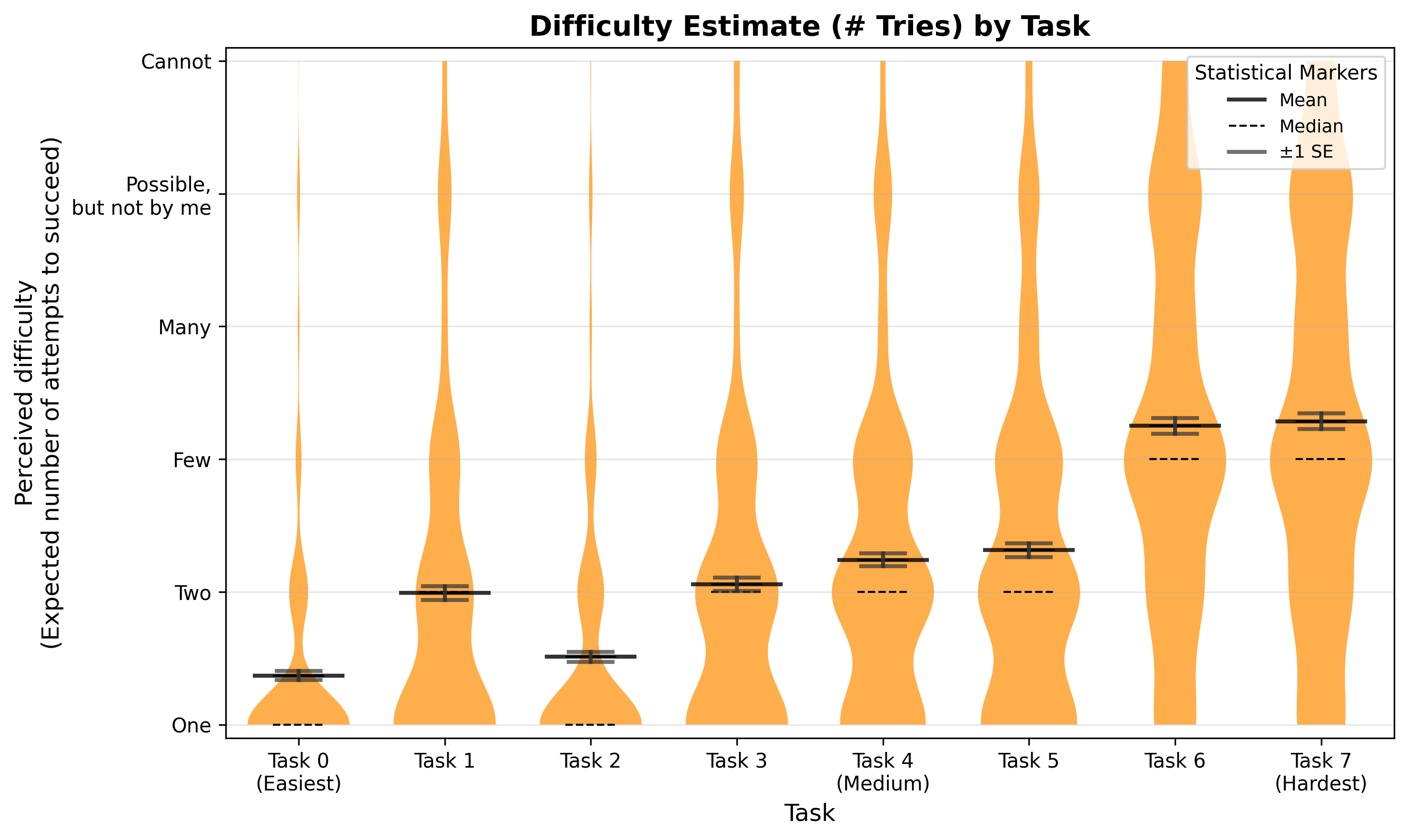}
    \caption{Participants were asked how many attempts they believed it would take to successfully generate the diagram in each Phase 2 task. Each task had more nodes or more complex connections than lower-number diagrams. In particular, Task 2 had one more node than Task 1, but Task 2 was sequential while Task 1 was cyclic, which could explain why Task 2 was viewed as easier. The task difficulties were selected in order to have two easier tasks, four tasks of middling difficulty, and two harder tasks. Tasks were displayed in randomized order.}
    \label{fig:perceived-difficulty}
\end{figure*}

\Cref{fig:perceived-difficulty} shows how many attempts participants believed it would take to successfully generate each Phase 2 diagram, illustrating that perceived difficulty generally increased with diagram complexity.

\subsection{The Effect of Prior AI Consumption}
\label{sec:prior_AI}

\begin{figure*}[ht]
    \includegraphics[width=\linewidth]{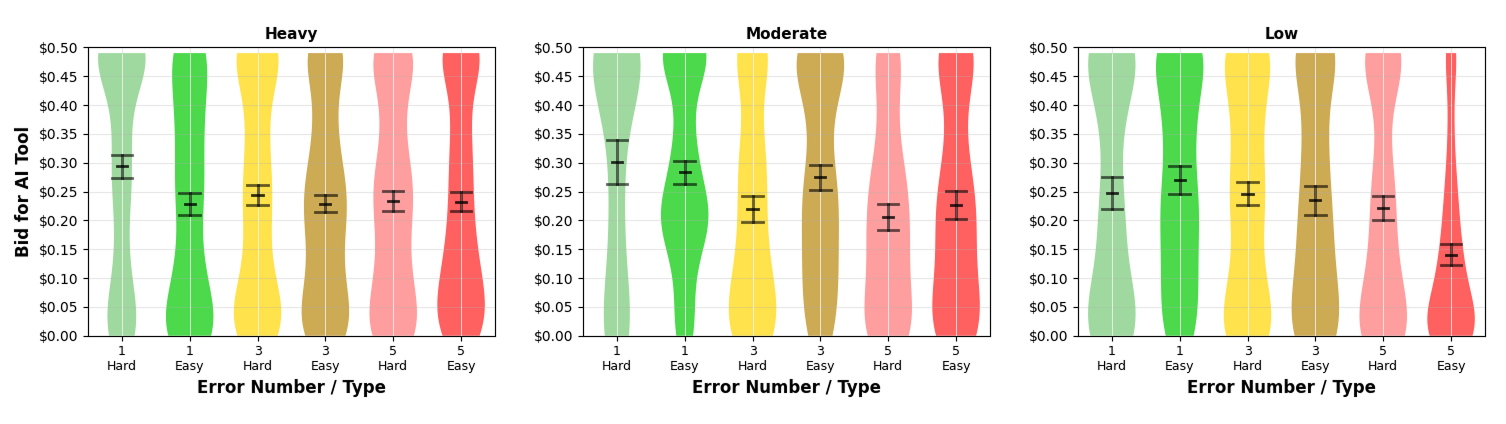}
    \caption{The empirical bid distribution for participants across the six experimental conditions, separated by prior AI consumption (heavy, moderate, and low). Error bars show standard errors.}
    \Description{Three violin plots for three separate groups of participants (heavy, moderate, and low AI-content consumers), each with six violin distributions, one for each experimental condition.}
    \label{fig:results-ai-consumption}
\end{figure*}

We found a potential influence of participants' prior consumption of AI-related content on reliance. This is based on the following question: ``How often do you read or watch AI-related stories, movies, TV shows, comics, news, product descriptions, conference papers, journal papers, blogs, or other materials?'' Participants answered by selecting one of the following choices: Daily, Weekly, Monthly, A few times a year, Rarely, or Never.

Consumption frequency categories were collapsed into heavy (Daily/Weekly), moderate (Monthly/A few times a year), and low (Rarely/Never) AI-content consumers. We present the bid distribution for each consumption category in \Cref{fig:results-ai-consumption}. We corrected for multiple comparisons by controlling for a 5\% false discovery rate, using the Benjamini-Hochberg procedure, within the 18 hypotheses that compare levels of AI-content consumption within an experimental condition and, separately, within the 45 hypotheses that compare experimental conditions within a level of AI-content consumption. Low AI-content consumers placed significantly higher bids for hard errors than easy errors in the 5-error condition (SE = \$0.02, $t$ = 5.265, $p_{\text{FDR}}$ < 0.001), whereas heavy and moderate users did not. The effect was significant after FDR correction ($p_{\text{FDR}}$ < 0.001). Within the 5-error/easy-task condition, low AI-content consumers placed significantly lower bids than heavy consumers with a difference of \$0.09 (SE = \$0.01, $t$ = 6.743, $p_{\text{FDR}}$ < 0.001) and moderate consumers with a difference of \$0.09 (SE = \$0.02, $t$ = 5.341, $p_{\text{FDR}}$ < 0.001).

\end{document}